\documentclass[12pt]{iopart}

\usepackage[hidelinks]{hyperref}
\usepackage{float}
\usepackage{graphicx}
\graphicspath{{}}
\begin{document}

\title[Mobility restores the mechanism which supports cooperation]
{Mobility restores the mechanism which supports cooperation in the voluntary prisoner's dilemma game}

\author{Marcos Cardinot$^1$, Colm O'Riordan$^1$, Josephine Griffith$^1$ and Attila Szolnoki$^2$}
\address{$^1$ Discipline of Information Technology, National University of Ireland Galway, H91FYH2, Ireland}
\address{$^2$ Institute of Technical Physics and Materials Science, Centre for Energy Research, Hungarian Academy of Sciences, P.O. Box 49, H-1525 Budapest, Hungary}

\ead{marcos.cardinot@nuigalway.ie}

\begin{abstract}
It is generally believed that in a situation where individual and collective interests are in conflict, the availability of optional participation is a key mechanism to maintain cooperation. Surprisingly, this effect is sensitive to the use of microscopic dynamics and can easily be broken when agents make a fully rational decision during their strategy updates. In the framework of the celebrated prisoner's dilemma game, we show that this discrepancy can be fixed automatically if we leave the strict and frequently artifact condition of a fully occupied interaction graph, and allow agents to change not just their strategies but also their positions according to their success. In this way, a diluted graph where agents may move offers a natural and alternative way to handle artifacts arising from the application of specific and sometimes awkward microscopic rules.
\end{abstract}

\vspace{2pc}
\noindent{\it Keywords}: evolutionary game theory, evolution of cooperation, optional prisoner's dilemma, mobility

\section{Introduction}

Despite extensive research efforts, the evolution of cooperation remains a
puzzle in a wide range of domains ~\cite{Perc2017,Smith1982}. In this context,
two-strategy games such as the prisoner's dilemma (PD) game have been widely
studied for many years from different perspectives with
mechanisms~\cite{Nowak2006b} such as group selection~\cite{Perc2013} and
network reciprocity~\cite{Szabo2007,Lieberman2005,Nowak1992} investigated.
Traditionally, the agents' interactions in those games are compulsory, i.e.,
the agent has to opt between cooperation or defection, where the dilemma arises
because individual selfishness leads to a collective
disaster~\cite{Axelrod1981,Rapoport1965}.  However, in many real-world
scenarios, the agents' participation in the game is voluntary (optional).
Thus, in order to account for the concept of voluntary participation (abstention),
researchers have been exploring the voluntary prisoner's dilemma (VPD) game,
also known as the optional prisoner's dilemma game, which extends the PD to a
three-strategy game where agents can also choose to abstain from playing the
game~\cite{Szabo2002,Batali1995,Orbell1993}.  In particular, abstention has
attracted attention both for acting as a mechanism to support cooperation and
for promoting cyclic
behaviour~\cite{CardinotSciRep,Jia2018a,Jia2018b,Hao2017,Jeong2014,Hauert2003}.
The cyclic dominance behaviour is often studied within the bounds of the
rock-paper-scissors game, which, different to the VPD game, imposes the
cyclic dominance in the payoff
matrix~\cite{Szolnoki2014,CardinotAICS,Szolnoki2009,Szabo2004a,Szabo2004b,duong_jmb19}.

In addition to the discussion about the game strategies, studies concerning
agent mobility are also of interest because, in many real ecological systems,
individuals are usually on the move to improve their
performance~\cite{Reichenbach2007}. In this sense, research has shown that in a
spatial environment, mobility and percolation thresholds have a critical impact
on the sustenance of biodiversity in
nature~\cite{Szolnoki2014,Yang2019,Yang2014,Chen2012,Wang2012a,Wang2012b,Liu2010}.
Interestingly, despite a large number of papers discussing the effects of
mobility in the prisoner's
dilemma~\cite{Vainstein2007,Vainstein2014,Sicardi2009,Zhu2013,Antonioni2014,Tomassini2015},
the rock-paper-scissors~\cite{Reichenbach2007,Szolnoki2016,Wang2010} and the
optional public goods games~\cite{Valverde2017,Zhong2013,Xia2012}, the impact of
mobility in the context of the VPD game is still almost unknown. Indeed, some
effort has also been made to explore contingent movement strategies modelling
the so-called ``win-stay, lose-move'' rule, which, as also argued by Szab{\'o}
and F{\'a}th~\cite{Szabo2007}, might capture the concept of abstention in the
sense that agents abstain by moving away from their
opponents~\cite{Aktipis2004,Hamilton2005,meloni_pre09,cardillo_pre12}. Although
this is a valid way to account for voluntary participation, we highlight that
in many scenarios there must be a cost (payoff) associated with the act of not
playing the game, i.e., abstention defined in terms of the set of game
strategies rather than the movement strategies. In other words, defining
abstention as a strategy rather than a movement ensures that all agents have
the right to abstain from a game interaction, independently of having a way to
walk away (space permitting) or not.

Despite the very recent introduction of the VPD game in a diluted network with
a purely random mobility scenario~\cite{Canova2018}, many questions regarding
the impact of mobility, in both the sustenance of biodiversity and the
potential for widespread cooperation, remain unanswered. For instance, given
the recent advances in the understanding of coevolutionary
models~\cite{CardinotPhysicaA,Szolnoki2018,CardinotECTA,Huang2015,Perc2010,zhang_yl_pone18,liu_rr_amc19},
what happens to the population when considering agent mobility in a
coevolutionary fashion?  Thus, without loss of generality, this research
introduces the VPD game with a coevolutionary model where not only the agents'
strategies but also their movement is subject to the evolutionary process,
which provides a more realistic representation of mobility within the domain of
voluntary/optional participation.

Furthermore, we investigate the foundations of the emergence of cyclic
dominance for the VPD game in both the fully populated (without mobility) and
diluted networks. We discuss that the emergence of the cyclic dominance
behaviour, which is commonly associated with the VPD game, is very sensitive to
the chosen imitation rule. Results show that when using other imitation rules,
the cyclic dominance can be broken easily, but this difference diminishes when
we use a more general diluted model where mobility can repair the missing chain
that is necessary to support cyclic dominance.

The remainder of the paper is organised as follows. Section~\ref{sec:methods}
describes the model and the experimental settings. Section~\ref{sec:results}
presents the results of the extensive Monte Carlo simulations, which allow us
to unveil the reason why mobility and optionality favour cooperation and cyclic
dominance. Finally, Section~\ref{sec:discussion} outlines the main conclusions.

\section{Methods}
\label{sec:methods}

In order to account for the features of the concept of voluntary participation
(abstention) and agent mobility, we consider a set of $N$ rational agents
playing the voluntary prisoner's dilemma game (also known as the optional
prisoner's dilemma game) on a $M \times M$ diluted square lattice network with
von Neumann neighbourhood and periodic boundary conditions, i.e., a toroid
where sites are either empty or occupied by an agent. In this way, to describe
the lattice occupation, we define the lattice's density as $\rho=N/M^2$
($0<\rho<1$), where $\rho=1$ means that the lattice is fully populated.

In the voluntary prisoner's dilemma (VPD) game, agents can be designated as a
cooperator ($C$), defector ($D$) or abstainer ($A$). Considering a pairwise
interaction, the payoffs are defined as follows: $D$ gets $P=0$ for mutual
defection, $C$ gets $R=1$ for mutual cooperation, $T=b$ for defection against a
cooperator, and $S=0$ for cooperation against a defector. Regardless of whether
one or two agents abstain, both agents get the loner's payoff $L=\sigma$, where
${R>L>P}$. Note that we adopt a weak version of the game, where ${T>R>L>P\ge
S}$ maintains the nature of the dilemma~\cite{Nowak1992,Szabo2002}.

We consider a randomly initialized population in which ${N/3}$ of each
strategy ($C$, $D$ and $A$) is distributed at random in the network.
Following the standard procedures of an asynchronous Monte Carlo (MC)
simulation in this context~\cite{Evoplex,Hauert2002}, at each MC time step,
each agent ($x$) is selected once on average to update its strategy and
position immediately. Thus, in one MC step, $N$ agents are randomly chosen to
perform the subsequent procedures: if the agent $x$ has no neighbours, it moves
to one of the four nearest empty sites (von Neumann neighbourhood) at random;
otherwise, the agent $x$ accumulates the utility $U_x$ by playing the VPD game
with all its nearest active (non-empty) neighbours ($\Omega_x$), selects one of
them at random (i.e., the agent $y$, which also acquires its utility $U_y$),
and considers copying its strategy with a probability given by the Fermi-Dirac
distribution function:
\begin{equation}
W = \frac{1}{1 + e^{(U_x-U_y)/K}}
\label{eq:fermi}
\end{equation}
where $K=0.1$ characterizes the amplitude noise to allow irrational
decisions~\cite{Szabo2007,Szabo2009}. In this research, we also consider the
scenario in which agents do not make irrational choices in the strategy
updating process (Equation~\ref{eq:fermi}), i.e., the agent $x$ only considers
copying $y$ if $U_y>U_x$.

After the agent $x$ updates its strategy, $U_x$ is recalculated, and $x$
considers moving to a random empty site (if any) in its neighbourhood with
probability:
\begin{equation}
W = \frac{1}{1 + e^{(u_x-v_x)/K}}
\label{eq:update}
\end{equation}
where $K=0.1$, $u_x=U_x/k_x$ is the agent $x$'s average utility, $k_x$ is the
number of active neighbours in $x$'s neighbourhood, and ${v_x=(u_x+\sum_{y \in
\Omega_x}{u_y})/(k_x+1)}$ is the average utility of $x$'s neighbourhood
including itself.  Thus, the agents that are performing worse (better) than
their neighbours have more (less) incentive to move.

Note that to make this research comparable with previous works, we consider the
absolute payoff during the strategy imitation process (Equation~\ref{eq:fermi}).
Nevertheless, it is noteworthy that our key results remain unchanged
qualitatively if we apply a degree-normalized payoff in this function.
However, in the case of mobility, the application of an absolute payoff in
Equation~\ref{eq:update} would cause an artifact effect. More precisely, it
would result in the erosion of a cooperative cluster because agents at the
periphery, who have fewer neighbours, would always be unsatisfied and move,
i.e., the mentioned cluster would shrink gradually.

In order to avoid finite size effects, results are obtained for different
network sizes, ranging from $M=200$ to $M=1000$. Simulations are run for a
sufficiently long relaxation time ($10^5$ or $10^6$ MC steps), where the final
level of each strategy is obtained by averaging the last $10^4$ MC steps.

\section{Results}
\label{sec:results}

In this section, we present some of the relevant experimental results obtained
when simulating a population of agents playing the voluntary prisoner's dilemma
(VPD) game on diluted square lattice networks, i.e., a coevolutionary model
where not only the agents' strategies but also their positions evolve over time.
Firstly, we consider the case in which the population is fully populated, i.e.,
density $\rho=1$, and we demonstrate that the emergence of cyclic dominance in
the VPD game is sensitive to the chosen dynamical rule because by using other
imitation rules the cyclic dominance can be broken easily. Secondly, we
investigate the case in which $\rho<1$ (diluted network), where we show that
mobility and dilution can repair the mechanisms necessary for supporting cyclic
dominance. Thirdly, we further investigate the micro-level evolutionary dynamics
for a diluted network both with and without mobility.

\subsection{Fully populated network\texorpdfstring{ ($\rho=1$)}{}: fragile cyclic dominance}
\label{sec:rho1}

\begin{figure}[htb!]
\centering
\includegraphics[width=0.55\linewidth]{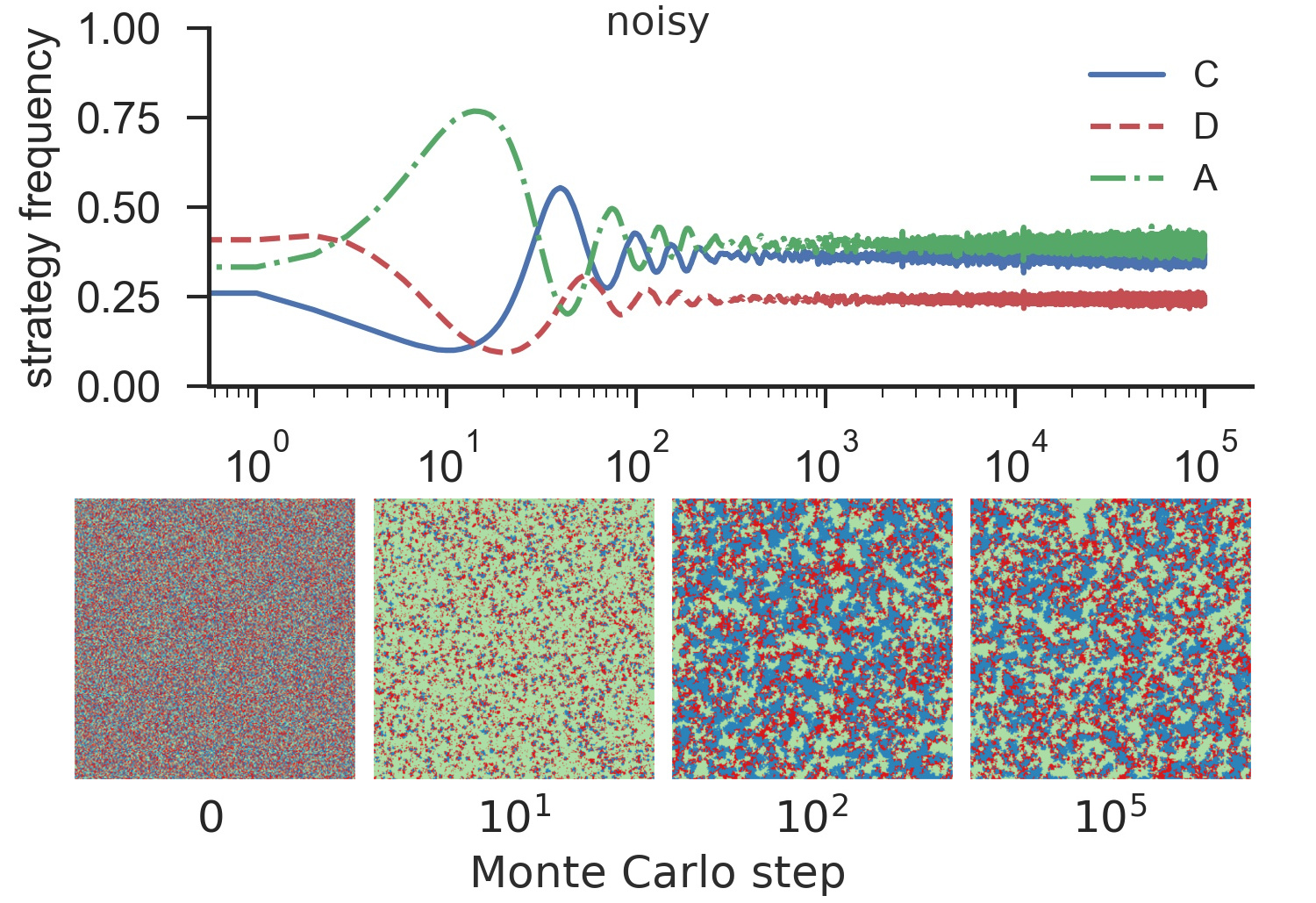}
\caption{Time course of the average frequency of cooperation (blue), defection (red) and abstention (green) for a fully populated network (density $\rho=1$) with $N=400^2$ agents, temptation to defect $b=1.4$, and loner's payoff $\sigma=0.5$ (top panel). Typical evolution of spatial distribution of the strategies (bottom panel). Results are obtained for the case in which agents are allowed to make irrational decisions, i.e., applies the noisy Fermi-Dirac imitation rule (Eq.~\ref{eq:fermi}) for any value of $U_y$ and $U_x$.}
\label{fig:fermi}
\end{figure}

In order to validate our coevolutionary model and provide grounds to explore
the effects of mobility on a diluted square lattice, we start by investigating
how the population evolves when there is no space for the agents to move.
Figure~\ref{fig:fermi} (upper panel) features the time course of the average
frequency of each pure strategy, i.e., cooperation, defection and abstention,
for a density $\rho=1$, temptation to defect $b=1.4$, and the loner's payoff
$\sigma=0.5$. The lower panel of Figure~\ref{fig:fermi} shows the typical
spatial patterns of the strategies at different Monte Carlo steps. Note that as
$\rho=1$, the model collapses to the traditional and well-known scenario in
which only the strategies evolve. As expected, the results are qualitatively
the same as those reported in previous studies~\cite{Szabo2002,Canova2018}.
In this case the three strategies coexist because of the emergence of cyclic
dominance behaviour where defectors beat cooperators, cooperators beat
abstainers, and abstainers beat
defectors~\cite{Szolnoki2014,CardinotAICS,hauert_s02}.

\begin{figure}[htb!]
\centering
\includegraphics[width=0.55\linewidth]{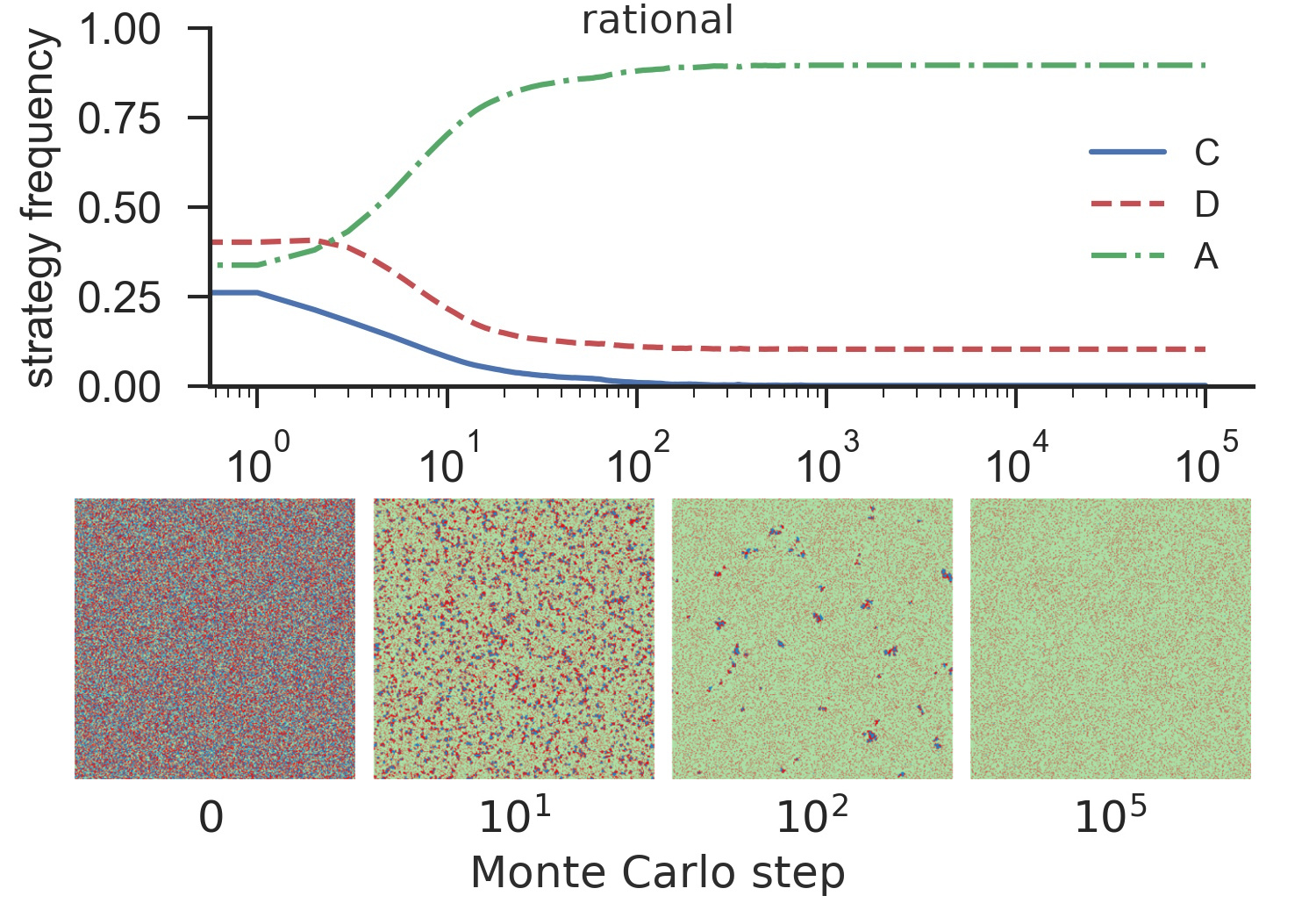}
\caption{Time course of the average frequency of cooperation (blue), defection (red) and abstention (green) for a fully populated network (density $\rho=1$) with $N=400^2$ agents, temptation to defect $b=1.4$, and loner's payoff $\sigma=0.5$ (top panel). Typical evolution of spatial distribution of the strategies (bottom panel). Results are obtained for the case in which agents make rational decisions only, i.e., applies the Fermi-Dirac imitation rule (Eq.~\ref{eq:fermi}) if and only if $U_y>U_x$. As compared to Fig.~\ref{fig:fermi}, note that cyclic dominance is quickly broken and cooperators die out soon because of the slight change in the imitation rule.}
\label{fig:posFermi}
\end{figure}

To gain deeper insights into the mechanisms which underlie the cyclic dominance
behaviour in the context of a spatial voluntary prisoner's dilemma game, we
perform the same experiments as above but for the case in which an agent ($x$)
only considers copying the opponent's strategy if the opponent ($y$) is
performing better than itself, i.e., applies the Fermi-Dirac distribution
function (Equation~\ref{eq:fermi}) if and only if the utility of $y$ is greater
than the utility of $x$, ${U_y>U_x}$.  Interestingly, Figure~\ref{fig:posFermi}
shows that when this simple modification in microscopic dynamics is imposed,
the cyclic dominance behaviour is broken and the population converges to a
frozen state where only defection and abstention are present, but the cooperator
strategy becomes extinct. Note that the idea of employing different imitation
rules such as Equation~\ref{eq:fermi} for both rational and irrational
decisions have been systematically investigated in previous studies for
two-strategy games~\cite{Antonioni2014,Nowak2009,roca_plr09}, and it is
well-known that different imitation rules, as well as the adoption of different
values of $K$ (amplitude noise) in the Fermi-Dirac rule may affect the outcome
\cite{Szabo2009}. However, there is an unexplored gap in the literature
regarding the possible consequences of the adoption of the Fermi-Dirac rule in
the context of the VPD game, and our results suggest that the cyclic behaviour
commonly associated with the VPD game may be related to the use of this function,
which also supports strategy change when the utility values are equal.

\begin{figure}[htb]
\centering
\includegraphics[width=.85\linewidth]{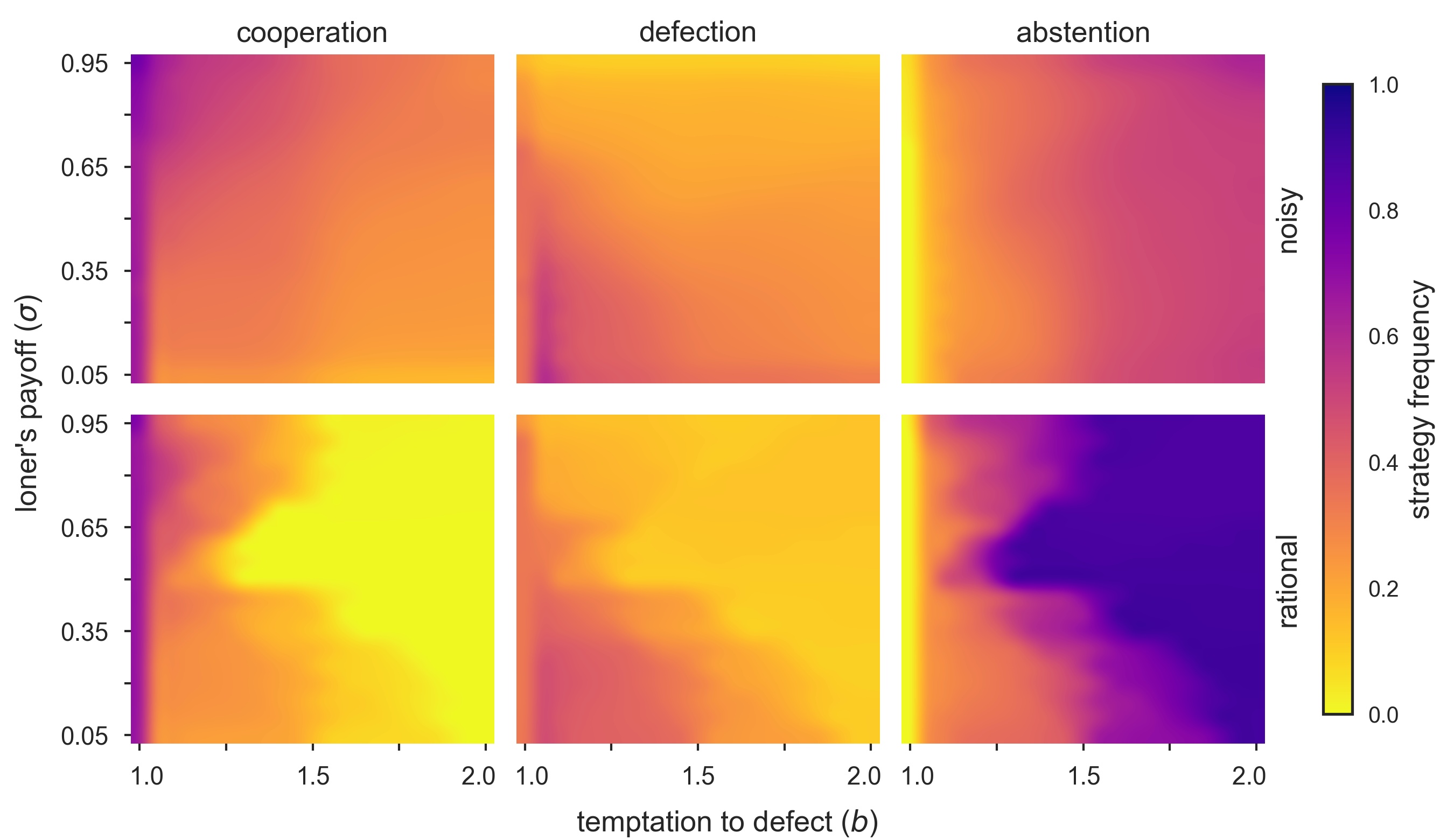}
    \caption{Heat maps of the average frequency of the strategies in the full $b-\sigma$ plane at the stationary state for a fully populated network (density $\rho=1$). Results for the noisy imitation rule on the top, and for exclusive rational imitation rule on the bottom row. Note that while cyclic dominance is maintained for almost any combination of $b$ and $\sigma$ values on the top, this behaviour is easily broken when changing the imitation rule to ``only copy if the opponent is performing better'' (bottom). In the latter case abstention becomes the dominant strategy for most $b-\sigma$ pairs.
    }
\label{fig:heatmaps_rho1}
\end{figure}

Figure~\ref{fig:heatmaps_rho1} depicts the average frequency of the three
strategies ($C$, $D$ and $A$) in the full $b-\sigma$ plane when agents are
allowed to make irrational (top panels) and rational (bottom panels) decisions.
Note that while cyclic dominance is maintained for almost any combinations of
$b$ and $\sigma$ values in the traditional case (top), the same does not occur
when the imitation rule is slightly changed (bottom). Thus, contrary to
previous observations, our results highlight that the use of noisy imitation,
dictated by Equation~\ref{eq:fermi}, is an essential condition for promoting
cyclic behaviour in the context of the VPD game. The reason for this
discrepancy can be summarized as follows:
\begin{itemize}
    \item Considering a random initial population (see the early MC steps in
        Figures~\ref{fig:fermi} and~\ref{fig:posFermi}), the typical trajectory
        predicts the advantage of defectors which is then followed by the rise
        of abstainers or both cases.

    \item Next, checking (or not) for the $U_y>U_x$ condition can be decisive
        to allow (or not) the subsequent rise of cooperators, which in turn
        supports the cyclic dominance phenomenon seen in Figure~\ref{fig:fermi}.

    \item At a micro level, if one cooperator/defector ($x$) is mostly
        surrounded by abstainers ($y$), its utility $U_x$ will be mostly equal
        to $U_y$. Remember that in the voluntary prisoner's dilemma game, if one
        or two agents abstain ($A$), both will get the same loner's payoff
        $\sigma$, i.e., for any pair of strategies $CA$, $AC$, $DA$, $AD$, $AA$ both
        agents get an identical $\sigma$ value.
\end{itemize}
Thus, if we impose the $U_y>U_x$ condition, as the utilities of $x$ and $y$ are
the same, the population is not able to curb the spreading of abstainers, which
consequently produces the pattern observed in Figure~\ref{fig:posFermi}, i.e.,
a few isolated defectors stuck in a sea of abstainers. Otherwise, if
Equation~\ref{eq:fermi} is applied for any value of $U_x-U_y$, as the number of
abstainers increase, $W$ will be approximately equal to $0.5$ for most agents,
which is one of the main mechanisms to keep the three strategies alive as
observed in Figure~\ref{fig:fermi}.

\subsection{Diluted network\texorpdfstring{ ($\rho<1$)}{}: recovering cyclic dominance and promoting cooperation}

As we already argued, a fully occupied interaction graph seems to be a specific
rather than a generally valid real life situation, hence this section
discusses the coevolutionary cases for a diluted lattice network where not only
the strategies but also the agents' positions evolve over time.

\begin{figure}[htb!]
\centering
\includegraphics[width=0.87\linewidth]{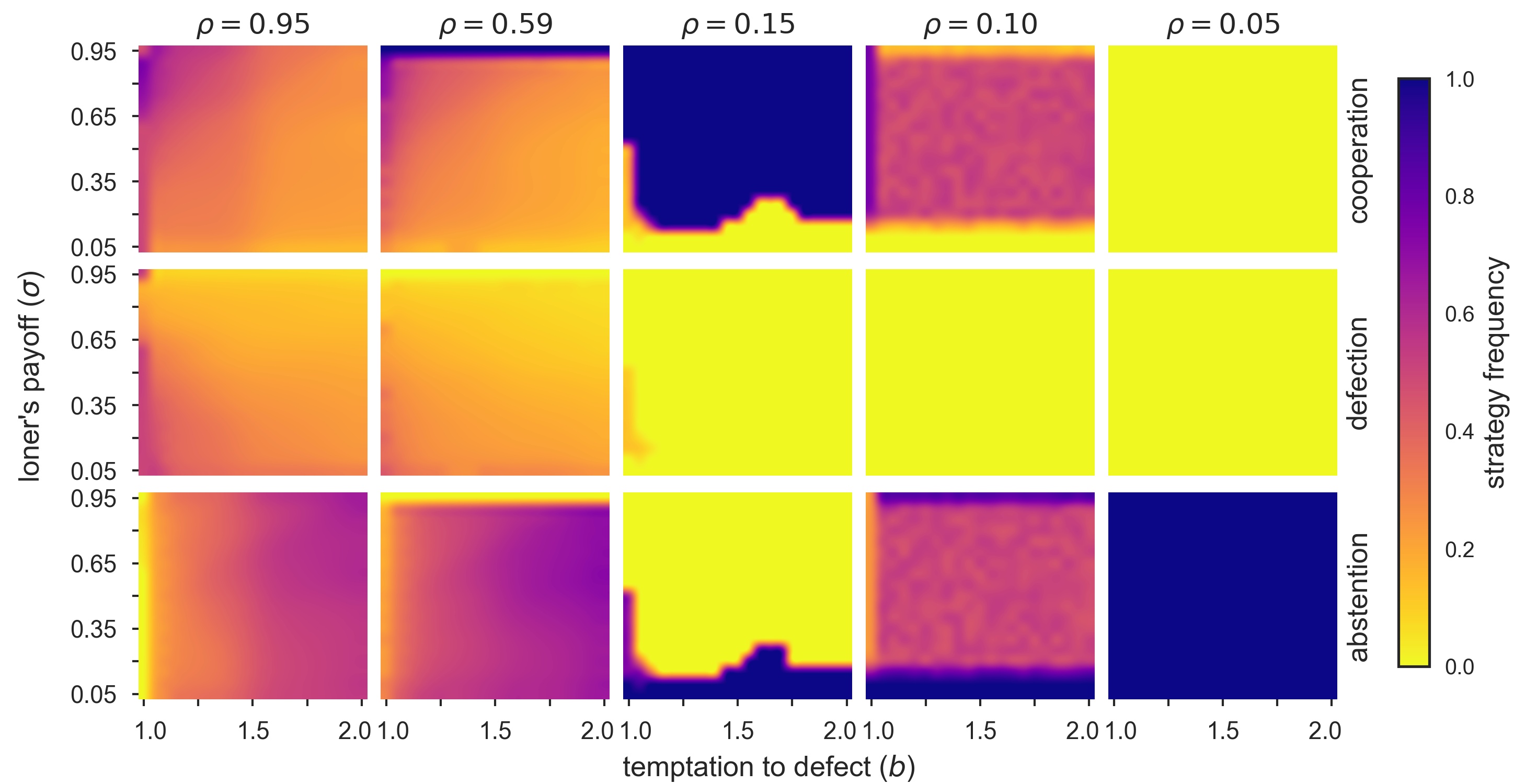}
\caption{Heat maps of the average frequency of cooperation (top), defection (middle) and abstention (bottom) in the full $b-\sigma$ plane at the stationary state for a diluted network. All results are obtained for the noisy imitation rule. Note that at $\rho=0.10$ global concentration, the average frequencies of $C$ and $A$ are approximately $0.5$ as a result of a bistable destination of the evolutionary process, where the population either converges to full dominance of cooperators or abstainers.}
\label{fig:heatmaps_fermi}
\end{figure}

At a macro-level, we start by analysing the influence of the density $\rho$ on
the evolutionary process for the noisy Equation~\ref{eq:fermi} (i.e., agents
are allowed to make irrational decisions) after a sufficiently long relaxation
time. In line with previous research for two-strategy games such as the
prisoner's dilemma game~\cite{Vainstein2007,Sicardi2009,Helbing2008},
experiments with our coevolutionary model reveal that mobility and dilution
also play a key role in promoting cooperation in the VPD game.
Figure~\ref{fig:heatmaps_fermi} shows the average frequency of the three
strategies in the full $b-\sigma$ plane for some representative densities. As
compared to the traditional case ($\rho=1.0$ regime i.e.,
Figure~\ref{fig:heatmaps_rho1} top), we observe that the cyclic dominance
behaviour still emerges for most $b-\sigma$ settings for $\rho\ge0.59$.
Interestingly, results show that scenarios of full cooperation arise
monotonously when $\rho<0.59$, i.e., the more diluted the network is, the
easier it is for cooperators to dominate the population. However, when the density
is too low ($\rho<0.10$) the cooperators become too vulnerable to invasion by
abstainers due to the increasing difficulty of forming clusters.  Also,
experiments show that $0.10\ge\rho>0.05$ quickly produces very unstable $C+A$
states which either converge to full $C$ or full $A$. Notably, this behavior
cannot be seen directly from the heat map because the average of full $C$
and full $A$ destinations results in around $0.5$ density for both strategies.
The latter may also suggest a coexistence of these strategies, but as we
stressed, not in the present case because either $C$ or $A$ prevails at these
global concentration values. Furthermore, when $\rho\le0.05$ cooperators always
die out and abstainers dominate in all scenarios.

Note that the percolation threshold ($\rho_p$) for this square lattice network
with von Neumann neighbourhood is approximately equal to
$0.59$~\cite{Malarz2005,Stauffer1994}. Thus, this result is of particular
interest because cooperation is favoured when the density is below the
percolation threshold, which is known to be an adverse situation for maintaining cooperation
\cite{Yang2014,Wang2012a,Wang2012b}. Moreover, results in
Figure~\ref{fig:heatmaps_fermi} also highlight the importance of exploring the
outcomes of the VPD game across the whole loner's payoff ($\sigma$) spectrum,
and not only for a specific $\sigma=0.3$ value, as was used earlier
~\cite{Szabo2002,Canova2018}.

\begin{figure}[htb!]
\centering
\includegraphics[width=.87\linewidth]{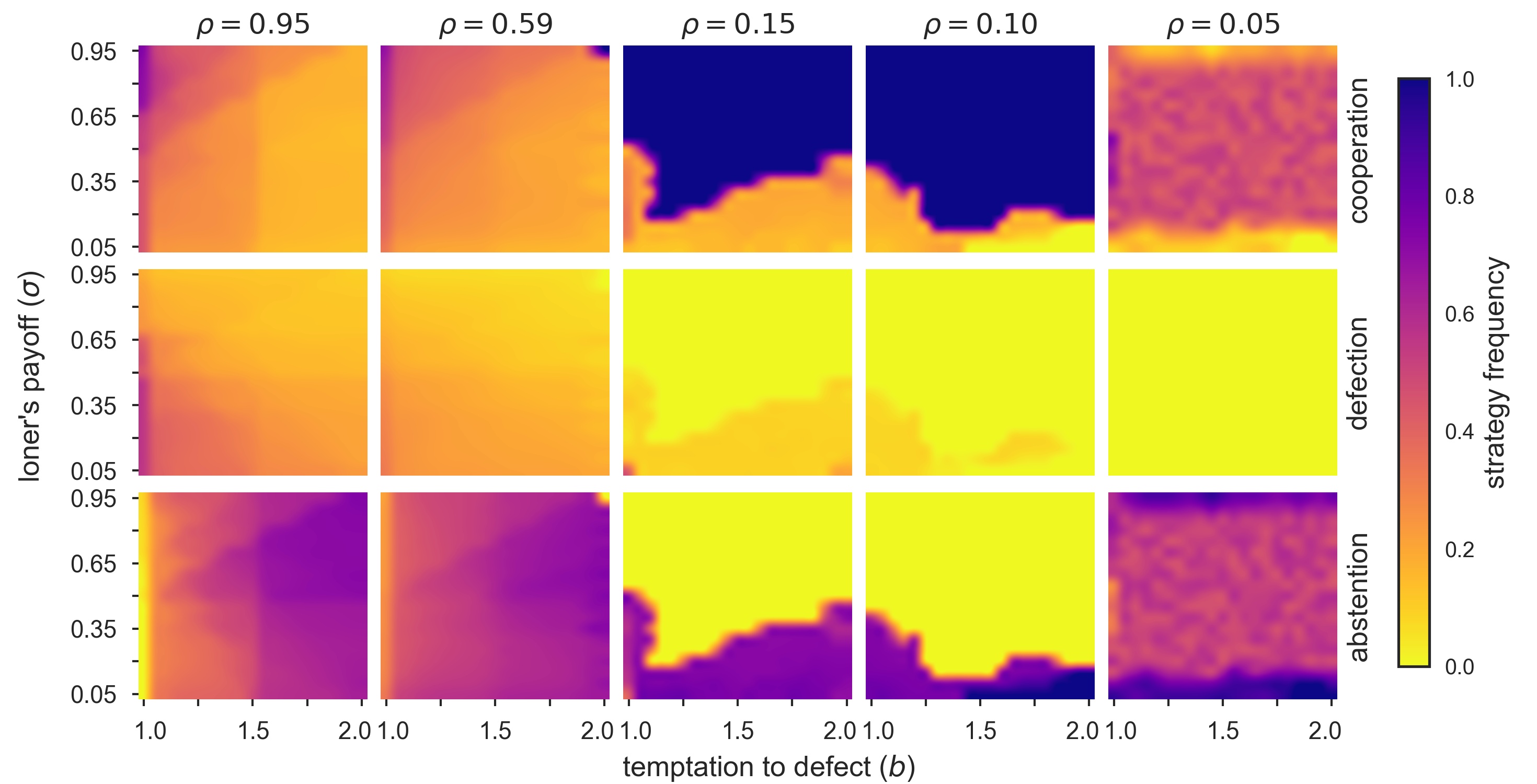}
\caption{Heat maps of the average frequency of cooperation (top), defection (middle) and abstention (bottom) in the whole $b-\sigma$ plane at the stationary state for a diluted network. All results are for the fully rational imitation rule.}
\label{fig:heatmaps_posFermi}
\end{figure}

Considering the discrepancy observed in Figure~\ref{fig:heatmaps_rho1} for
$\rho=1$, we now repeat the same experiments as above but for the case where an
agent only applies Equation~\ref{eq:fermi} if the opponent is performing better
than itself, i.e., the case of a fully rational imitation rule. Surprisingly,
Figure~\ref{fig:heatmaps_posFermi} shows that the previously observed
difference for both imitation rules diminishes when we consider a diluted
network ($\rho<1$) with mobile agents. More importantly, results show that when
$1>\rho>\rho_p$ the mechanisms which support cyclic dominance in the
traditional case (i.e., for the noisy Equation~\ref{eq:fermi} and $\rho=1$) are
recovered for a wide range of $b-\sigma$ scenarios. In fact, results for both
imitation rules and $\rho<1$ are qualitatively the same for most settings.
However, as seen in figures~\ref{fig:heatmaps_fermi}
and~\ref{fig:heatmaps_posFermi}, when the density is below the percolation
threshold $\rho<\rho_p$, it is possible to observe a small shift of
$\rho\approx0.05$ in the boundaries of the region in which full $C$ occurs. For
instance, results for $\rho=0.15$ in Figure~\ref{fig:heatmaps_fermi} are
similar to those when $\rho=0.10$ in Figure~\ref{fig:heatmaps_posFermi}.
Note that the bistable outcomes, where the population either converges to a full
$C$ or a full $A$ state, observed for $\rho\approx0.10$ in the first
case happens at $\rho\approx0.05$ in the later case.

\subsection{Micro-level analysis of the effects of dilution and mobility}

\begin{figure}[htb!]
\centering
\includegraphics[width=0.85\linewidth]{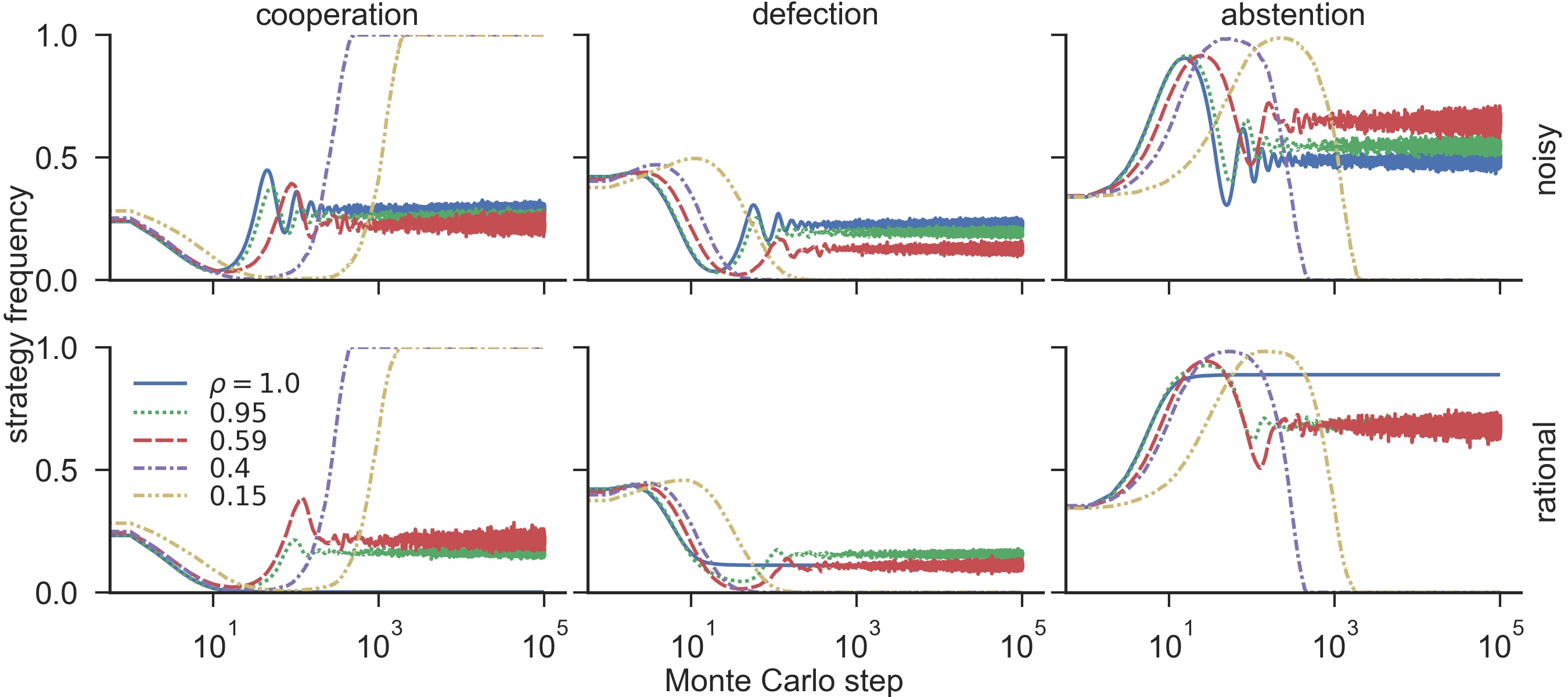}
\caption{Time course of the average frequency of cooperation (left), defection (center) and abstention (right) for different densities $\rho$, temptation to defect $b=1.65$, and loner's payoff $\sigma=0.55$. Results for the noisy imitation rule on the top, and the rational imitation rule on the bottom. Note that the difference between both rules diminishes when $\rho<1$.}
\label{fig:time}
\end{figure}

In order to further explore the aforementioned phenomena, we extend our
analysis of the evolutionary process to a micro perspective.
Figure~\ref{fig:time} shows the average time course of the three strategies for
a fixed temptation to defect $b=1.65$ and loner's payoff $\sigma=0.55$, which
is representative of the outcomes of other parameters as well. For this
scenario, when $\rho=1$, cyclic dominance is maintained for the traditional
case with the noisy imitation rule, but it is easily shattered when considering
a rational rule. However, the difference diminishes when $\rho<1$.

Results show that the profile of the curves for the initial $10^2$ MC steps are
very similar to scenarios which support cyclic dominance, i.e., an initial
drop followed by a quick recovery of the frequency of cooperators. This
phenomenon has also been observed in previous work for dynamic
networks~\cite{CardinotPhysicaA,Szolnoki2008a}, where it was discussed that
defectors are quickly dominated by abstainers, allowing a few clusters of
cooperators to remain in the population, then with the lack of defectors, those
cooperative clusters expand by invading the abstainers.  Note that it also
explains the reason that higher values of $\sigma$ are more beneficial in
promoting cooperation (as seen in
Figures~\ref{fig:heatmaps_rho1},~\ref{fig:heatmaps_fermi}
and~\ref{fig:heatmaps_posFermi}), i.e., abstainers have to be strong enough to
protect cooperators against invasion from defectors in the initial steps.
Moreover, Figure~\ref{fig:time} (right) shows a clear correlation between the
density $\rho$ and the speed of the initial inflation of abstention.

\begin{figure}[htb!]
\centering
\includegraphics[width=0.85\linewidth]{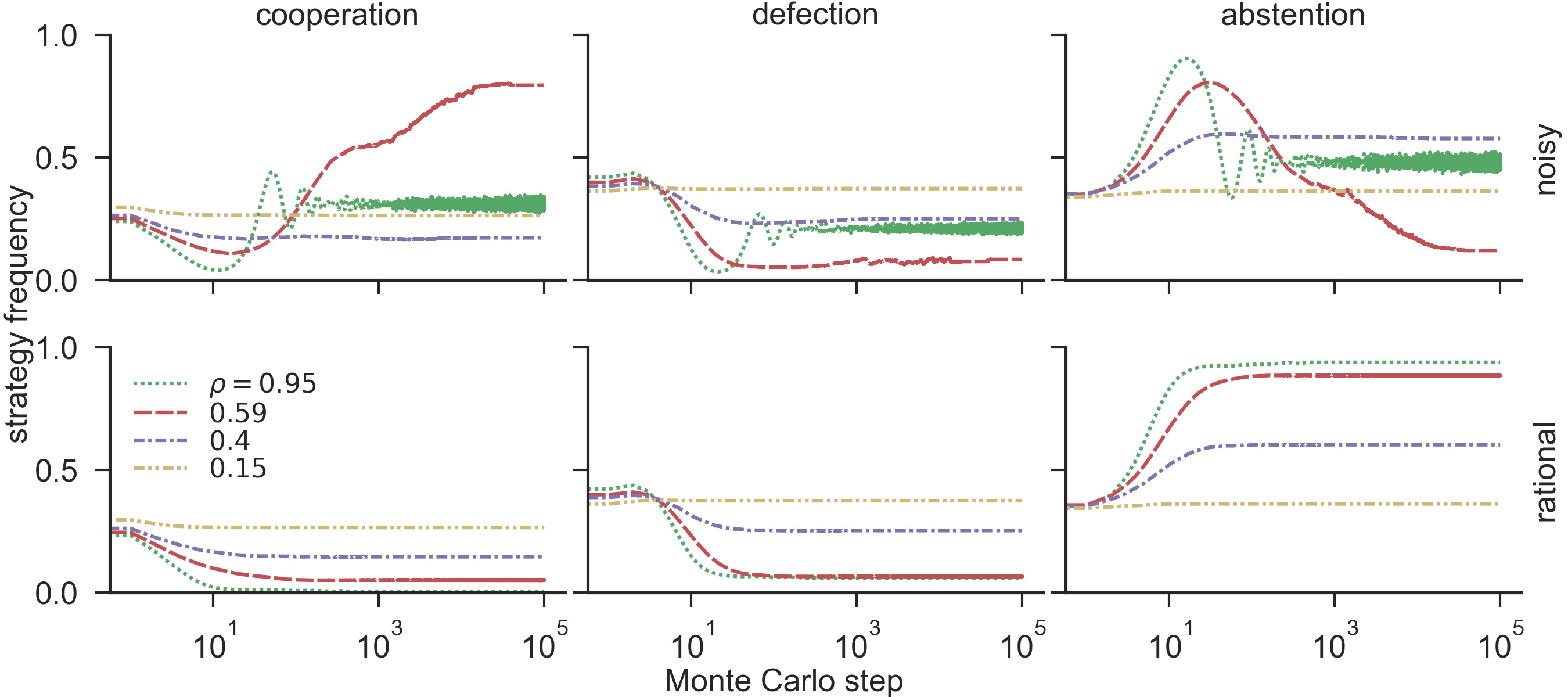}
\caption{Time course of the average frequency of cooperation (left), defection (center) and abstention (right) for different densities $\rho$, temptation to defect $b=1.65$, and loner's payoff $\sigma=0.55$. Results obtained for a diluted network \textit{without} mobility. The panels compare the noisy imitation rule (top) and the rational imitation rule (bottom).}
\label{fig:noMobility}
\end{figure}

In order to distinguish between the impact of mobility and dilution on the
emergence of cooperative behaviour and cyclic dominance, we have also
investigated the case in which the agents are not allowed to move. That is, the
same model described in Section~\ref{sec:methods}, but without the movement
updating process. As shown in Figure~\ref{fig:noMobility}, when $\rho\le\rho_p$
the frequency in which the agents change their strategies is extremely low,
i.e., the population quickly reaches a frozen pattern which is very dependent
on the initial configuration. Also, in line with preceding
research~\cite{Wang2012a,Wang2012b}, we observe that when considering the
traditional noisy imitation rule (Figure~\ref{fig:noMobility} top), dilution
alone can improve the level of cooperation, where the optimal value of $\rho$
is always above the percolation threshold ($1>\rho>\rho_p$). In another
perspective, the emergence of cyclic dominance behaviour is diminished when the
agents do not move (e.g., compare the top panels of the Figures~\ref{fig:time}
and~\ref{fig:noMobility}).

Interestingly, different phenomena occur when we consider the fully rational
imitation rule (Figure~\ref{fig:noMobility} bottom). Note that dilution alone
is not able to fix the evolutionary mechanisms which support either the
emergence of cyclic dominance and the evolution of cooperation. In other words,
results show that mobility plays a key role in diminishing the difference on
the outcomes of both imitation rules (as seen in Figure~\ref{fig:time} for
$\rho<1$). Moreover, it is noteworthy that mobility allows for the full
dominance of cooperation for lower values of $\rho$, as well as the robust
emergence of cyclic dominance for a wider range of scenarios.

\begin{figure}[htb!]
\centering
\includegraphics[width=0.6\linewidth]{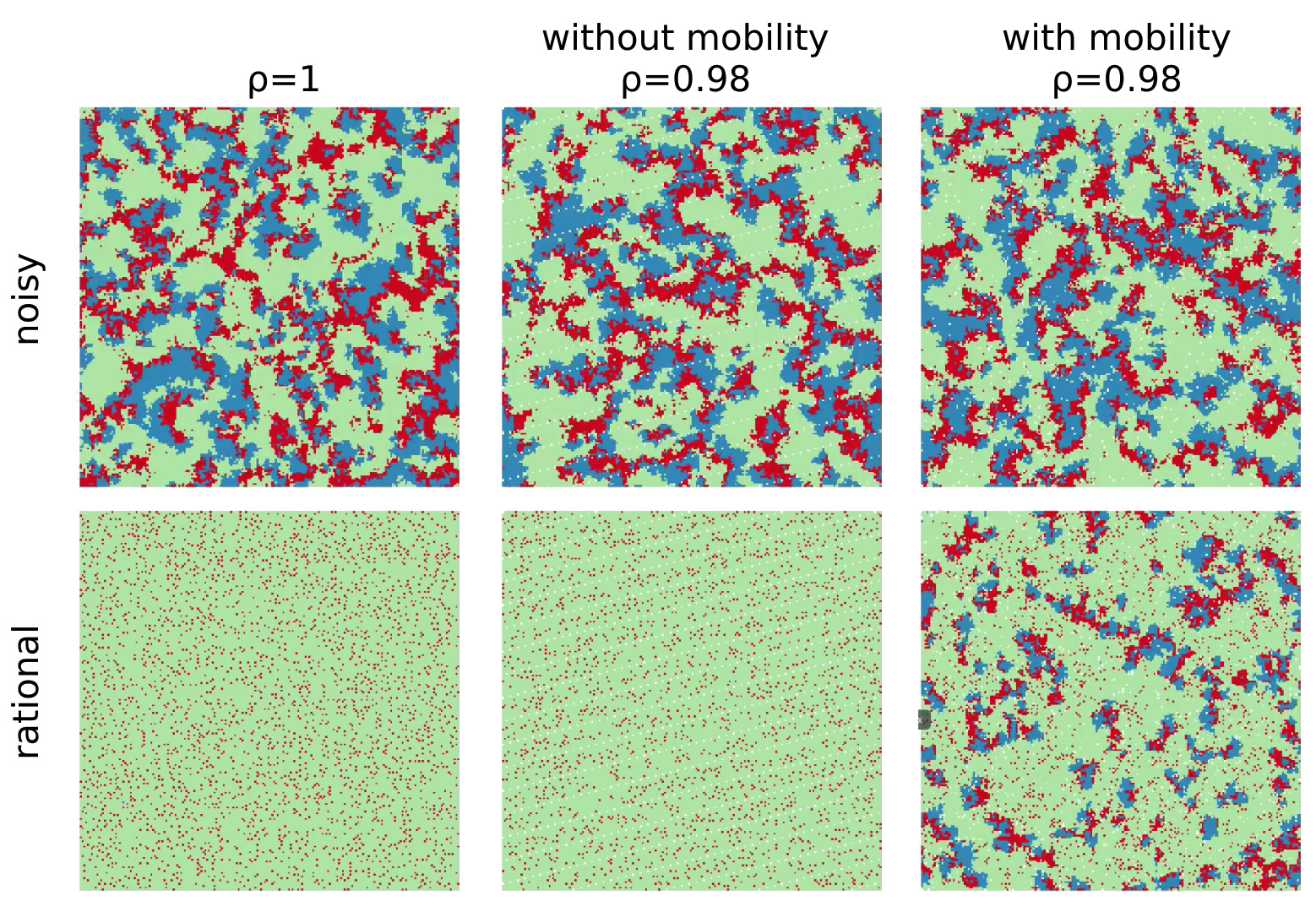}
\caption{Snapshots of the spatial evolution of cooperation (blue), defection (red) and abstention (green) in the final stationary state with different scenarios.
All simulations run for the same loner's payoff $\sigma=0.55$, and temptation to defect $b=1.65$. Results for the noisy imitation rule on the top, and the fully rational imitation rule on the bottom. An animation of the evolution of the strategies is provided (\url{https://doi.org/10.6084/m9.figshare.8038988}).}
\label{fig:animation}
\end{figure}

To advance the understanding of mobility and dilution in the context of the
VPD game, we also analyse the spatio-temporal dynamics of the strategies for
both the noisy and the rational imitation rules. Figure~\ref{fig:animation}
provides an animation for a prepared initial state where the strategies are
arranged in stripes. This prepared configuration allows us to separate
cooperators from defectors, making it easier to observe the mechanisms which are
responsible for breaking the cyclic chain where $A$ beats $D$, $D$ beats $C$
and $C$ beats $A$.  In summary, results show that the key difference between
the dynamical rules is that, when applying the fully rational rule, defectors
in the middle of abstainers do not have the incentive to become abstainers.
Hence, as discussed in Section~\ref{sec:rho1}, the rational rule produces
frozen $D+A$ states (as seen in Figure~\ref{fig:posFermi}) which cannot be
observed in the noisy Fermi-Dirac case. As a consequence, the isolated
defectors trapped in the sea of abstainers inhibit the formation of larger
cooperative clusters, which in turn breaks the cyclic chain. However, when
mobility is introduced for $\rho<1$, the $D+A$ states are not a stable phase
anymore. Here, there is a small stir which causes a random drift of defectors.
Consequently, when two defectors meet they become vulnerable against invasion
from abstainers. This process would lead to a homogeneous $A$ phase, but the
latter is sensitive to the attack of cooperators. In this way, abstainers are
now able to support the emergence of cooperation, which in turn restores the
mechanism to maintain the coexistence of all competing strategies.

Furthermore, regarding the phenomenon of cyclic dominance observed when
${\rho<1.0}$, although using a different scenario and methodology, our results
are compatible with previous research concerning mobility in the
rock-paper-scissors game, where it is discussed that mobility can jeopardise
cyclic dominance~\cite{Reichenbach2007,Jiang2011}. However, in the context of
the VPD game, the enhancement of cooperation for $\rho<\rho_p$ is
counter-intuitive because it diminishes the cooperators' ability to form larger
clusters~\cite{Canova2018}. Besides, results show that when the agents are
allowed to abstain, the population of mobile agents will never converge to full
defection. Finally, it is noteworthy that results also echo the findings of
previous research concerning the PD and VPD games on weighted
networks~\cite{CardinotPhysicaA,CardinotECTA,Huang2015}, i.e., a coevolutionary
model in which the link weights are also subject to evolution. In parallel, the
ability of avoiding interactions either by weakening the link weight or by
moving to another position acts as an important mechanism to strengthen
cooperators against exploitation.

\section{Discussion and Conclusions}
\label{sec:discussion}

This work investigates the role of mobility and dilution in a population of
agents playing the voluntary prisoner's dilemma (VPD) game, also known as the
optional prisoner's dilemma game, in a diluted square lattice network. We
propose a coevolutionary model where both the agents' strategy and position are
subject to evolution. In this model, in addition to the commonly applied
imitation rules for the strategies~\cite{Szabo2002}, we also adopt a mobility
rule in which agents who are performing worse (better) than their neighbours
have more (less) chance to move. Thus, without loss of simplicity, this
coevolutionary and asynchronous model is more realistic than the previous ones
which consider random mobility with synchronous updating
rules~\cite{Canova2018}.

Research in this domain has claimed that the addition of abstention in the
prisoner's dilemma game leads to a rock-paper-scissors type game, in which
cooperation dominates abstention, abstention dominates defection, and
defection, in turn, dominates cooperation, which describes the so-called cyclic
dominance behaviour~\cite{Szolnoki2014}. Interestingly, the present study shows
that, in the context of the traditional VPD game for a fully populated
network~\cite{Szabo2002}, the emergence of cyclic behaviour is biased by the
use of the Fermi-Dirac distribution function (sigmoid) in the strategy adoption
process. This sigmoid function is often employed to allow for irrational or
unjustified decisions where agents occasionally copy the strategy of a worse or
an equally performing
neighbour~\cite{Szabo2007,Szabo2009,Szabo2005,du_f_csf13,javarone_jsm16}. We
show that when agents make fully rational decisions such as only copying the
strategy of better performing neighbours, the outcome changes drastically,
making cyclic behaviour unsustainable in most cases. However, the present study
shows that the mechanism that supports cyclic behaviour is fixed when agents
are allowed to move due to a diluted interaction space.

In fact, the noisy strategy updating rule has been applied to avoid artifact or
frozen outcomes. However, in the present study we show that it is also possible
to avoid such frozen states in a more realistic way, where, for instance,
agents are allowed to move and change their connections over time.  Hence, a
deterministic rule can be as efficient as the noisy Fermi-Dirac function if we
assume a partly diluted system.  Furthermore, by means of robust and systematic
Monte Carlo simulations, results show that mobility plays a crucial role in
promoting cooperation in the VPD game for a wide range of values of the
temptation to defect $b$, and loner's payoff $\sigma$, including for scenarios
of high $b$ and density below the percolation threshold $\rho<\rho_p$, which
are known to be adverse for maintaining cooperative
behaviour~\cite{Yang2014,Wang2012a,Wang2012b}.

To conclude, this paper aims to bridge the gap between agent mobility and the
concept of voluntary/optional participation in social dilemmas. In addition, it
provides a novel perspective for understanding the foundations of cyclic
dominance behaviour in the context of the prisoner's dilemma game with
voluntary participation (VPD game). We hope this work can serve as a basis for
further research on the role of abstention to advance the understanding of the
evolution of cooperation in coevolutionary spatial games.

\section*{Acknowledgments}
This work was supported by the National Council for Scientific and Technological Development (CNPq-Brazil, Grant 234913/2014-2) and by the Hungarian National Research Fund (Grant K-120785).

\section*{References}
\bibliographystyle{iopart-num}
\providecommand{\newblock}{}

\end{document}